\documentclass[a4paper]{revtex4}
\usepackage[top=1in, bottom=1in, left=1in, right=1in]{geometry}
\usepackage{amsmath}
\usepackage{amssymb}
\usepackage{amsfonts}
\usepackage{graphicx}
\usepackage{dcolumn}
\usepackage[colorlinks,linkcolor=blue,citecolor=blue,urlcolor=blue]{hyperref}

\begin{document}

\title{Late-time acceleration with steep exponential potentials }
\author{M. Shahalam$^1$\thanks{E-mail address: shahalam@zjut.edu.cn},
 Weiqiang Yang$^2$\thanks{E-mail address: d11102004@163.com},
 R. Myrzakulov$^3$\thanks{rmyrzakulov@gmail.com},
 Anzhong Wang$^{1,4}$\thanks{Anzhong$\_$Wang@baylor.edu}}
\affiliation{$^{1}$Institute for Advanced Physics $\&$ Mathematics, Zhejiang University of Technology, Hangzhou, 310032, P. R. China\\
$^2$Department of Physics, Liaoning Normal University, Dalian, 116029, P. R. China\\
$^3$Eurasian International Center for Theoretical Physics, Department of General and Theoretical Physics, Eurasian National University, Astana, 010008, Kazakhstan\\
GCAP-CASPER, Department of Physics, Baylor University, Waco, TX, 76798-7316, USA }
\begin{abstract}
In this letter, we study the cosmological dynamics of steeper potential than exponential. Our analysis shows that a simple extension of an exponential potential allows to capture late-time cosmic acceleration and retain the tracker behavior. We also perform statefinder and $Om$ diagnostics to distinguish dark energy models among themselves and with $\Lambda$CDM. In addition, to put the observational constraints on the model parameters, we modify the publicly available CosmoMC code and use an integrated data base of baryon acoustic oscillation, latest Type Ia supernova from Joint Light Curves sample and the local Hubble constant value measured by the Hubble Space Telescope. 
\end{abstract}

\maketitle
\section{Introduction}
\label{sec:intro}
A large number of cosmological observations suggest that the present universe is undergoing a period of an accelerated expansion directly \cite{1,2,3} and indirectly \cite{4,5,6,7,8,9,10,11,12,13}. In the Einstein theory of gravity, dark energy (DE, an exotic fluid with large negative pressure) might be responsible for the current accelerated expansion of the universe \cite{14,15}. The simplest candidate of DE is the cosmological constant, and known as $\Lambda$CDM model. However, it is plagued with different theoretical problems, namely fine tuning and cosmic coincidence \cite{16,17,18,19,20}. Therefore, this is important to understand the nature of DE whether it is cosmological constant or it has dynamics.

Scalar fields play a key role in cosmology, and usually known as quintessence \cite{21,22,23,24,25}. The pressure could be negative for a slowly rolling scalar field if the potential energy is larger than the kinetic energy. A slowly varying scalar field rolls down a potential that might be responsible for explaining the late-time cosmology. To understand the applications of dynamical scalar fields, one has to study the characteristics of their potentials.

In case of quintessence, the scalar field models can be separated into two groups; thawing (slow-roll) and freezing (fast-roll) \cite{linder05}. Thawing models are very sensitive to the initial conditions whereas freezing models are independent for a wide range of initial conditions \cite{msa1,msa2}. In this paper, we shall focus on freezing models. Further, freezing models can be divided into two categories such as scaling and tracker \cite{wand,shinji}. A tracker model provides late-time acceleration while it is not possible in scaling models. To this effect, we consider steeper potential than exponential. The field energy density of a standard exponential potential does not evolve in the past due to a large Hubble damping. As time passes, it evolves and scales with the background at the present epoch, and remain so in future. Hence, in this case, we never get late-time acceleration. In case of steeper potential than exponential, the field energy density  freezes in the past due to increased value of Hubble damping. After sometimes, it evolves and scales with the background around the current epoch. Thereafter, it exits to the background and provides late-time acceleration at present epoch. The detail investigations of tracker solutions have been described in Ref. \cite{stein}.

The paper is organized as follows. Section \ref{sec:sfd} is devoted to scalar field dynamics and their evolution equations. In section \ref{sec:om}, we study statefinder and $Om$ diagnostics, and apply them to the underlying models. We put observational constraints on the model parameters by using joint analysis of baryon acoustic oscillation (BAO), latest Type Ia supernova (SNIa) from Joint Light Curves (JLA) sample and the local Hubble constant value measured by the Hubble Space Telescope (HST) through CosmoMC code in section \ref{sec:data}. Our results are summarized in section \ref{sec:conc}.
\section{Scalar field dynamics}
\label{sec:sfd}
\begin{figure}[tbp]
\begin{center}
\begin{tabular}{cc}
{\includegraphics[width=2.1in,height=2.1in,angle=0]{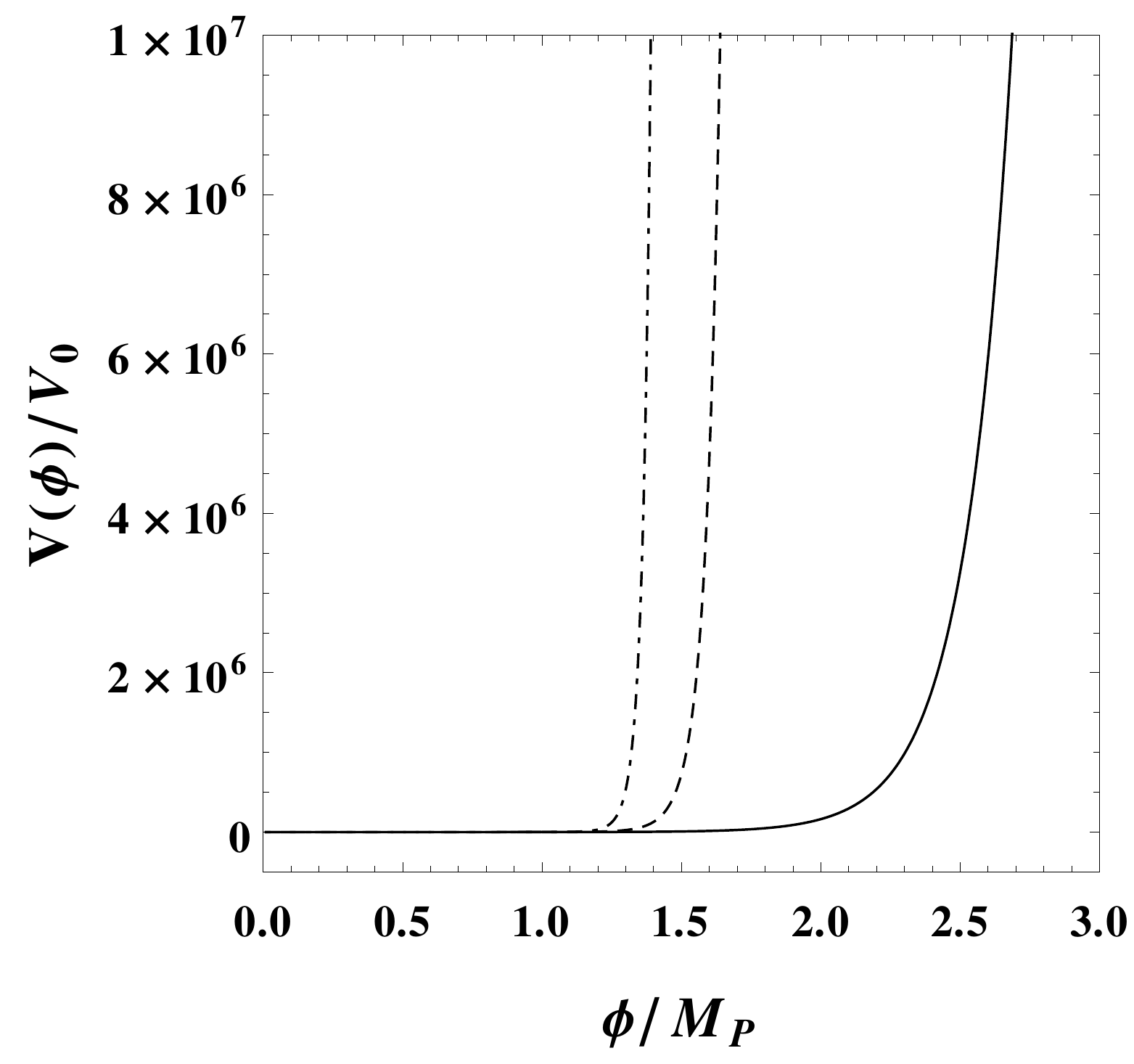}} &
{\includegraphics[width=2.1in,height=2.1in,angle=0]{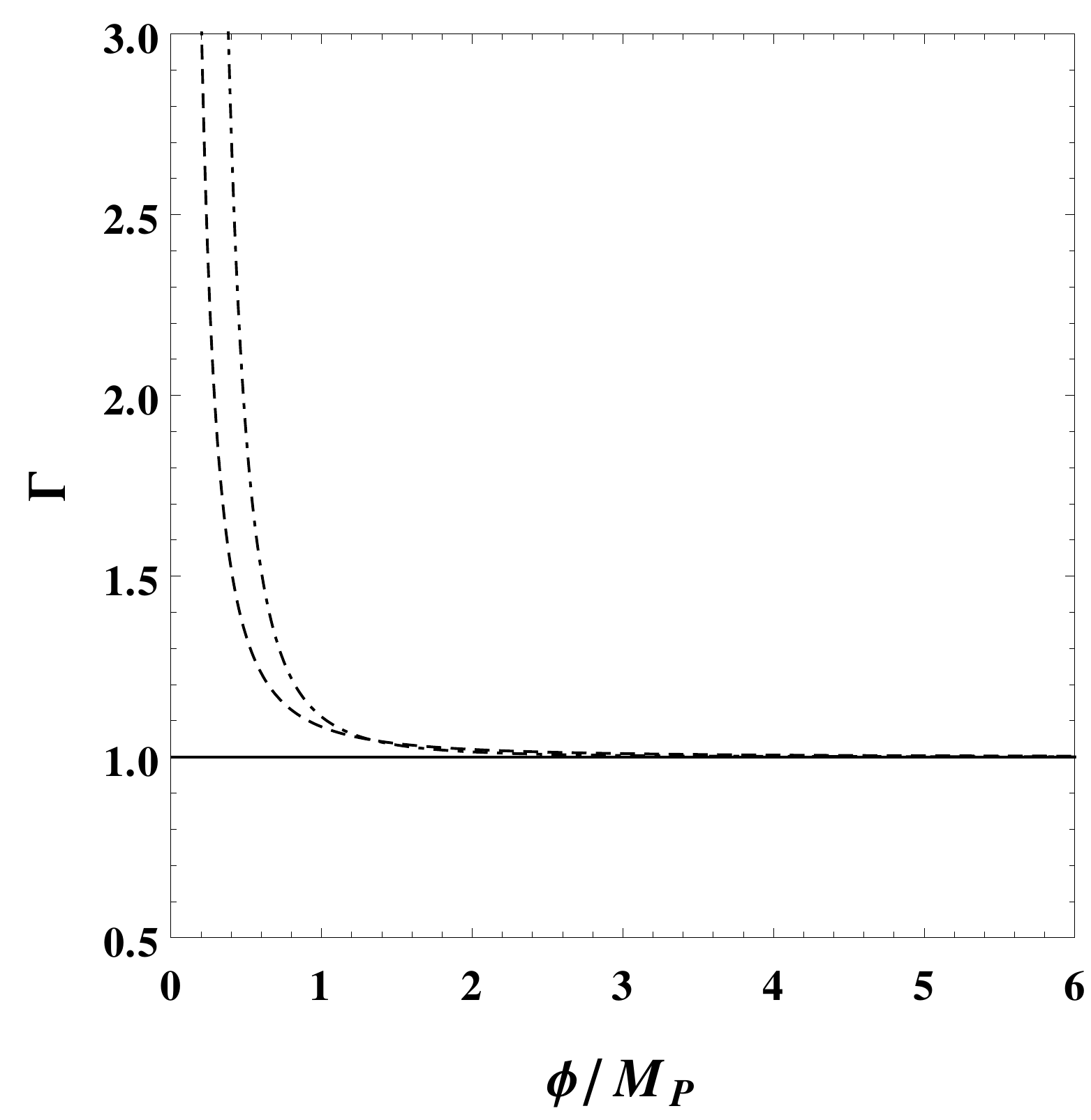}}
\end{tabular}
\end{center}
\caption{ This figure shows the evolution of potential (\ref{eq:pot}) and the function $\Gamma$ (\ref{eq:gamma}) versus field $\phi$. The figure is plotted for $n=1$ (solid line), 2 (dashed) and 3 (dot-dashed) with $\alpha=6$. For $n=1$, the behavior of $\Gamma$ is constant and remains unity throughout the evolution. For larger values of $\phi$, the behavior of $\Gamma$ is similar to the standard exponential potential irrespective of $n$, whereas for smaller values of $\phi$, it shows deviation from unity in case of $n=2$ and 3. }
\label{fig:v}
\end{figure}

In this section, we study the cosmological dynamics of standard and steeper exponential potentials. To this effect, we consider following form of the potential.
\begin{eqnarray}
V(\phi)=V_0 ~  e^{\alpha (\frac{\phi}{M_p})^n}
\label{eq:pot}
\end{eqnarray}
where $M_p$ is the reduced Planck mass, the parameter $V_0$ represents the dimension of $M_p^4$, $n$ denotes a number, and $\alpha$ is a dimensionless parameter. Here, we shall consider only positive values of $\alpha$ in order to get late-time cosmic acceleration. For negative values of $\alpha$, late-time acceleration is not possible, however, it can be obtained only when there is a nonminimal coupling between scalar field and neutrinos \cite{sami2017}. For $n=1$, potential (\ref{eq:pot}) reduces to the standard exponential potential. The evolution of the potential with respect to field for different values of $n$ is shown in the left panel of Fig. \ref{fig:v}.
\begin{figure}[tbp]
\begin{center}
\begin{tabular}{c}
{\includegraphics[width=2.1in,height=2.1in,angle=0]{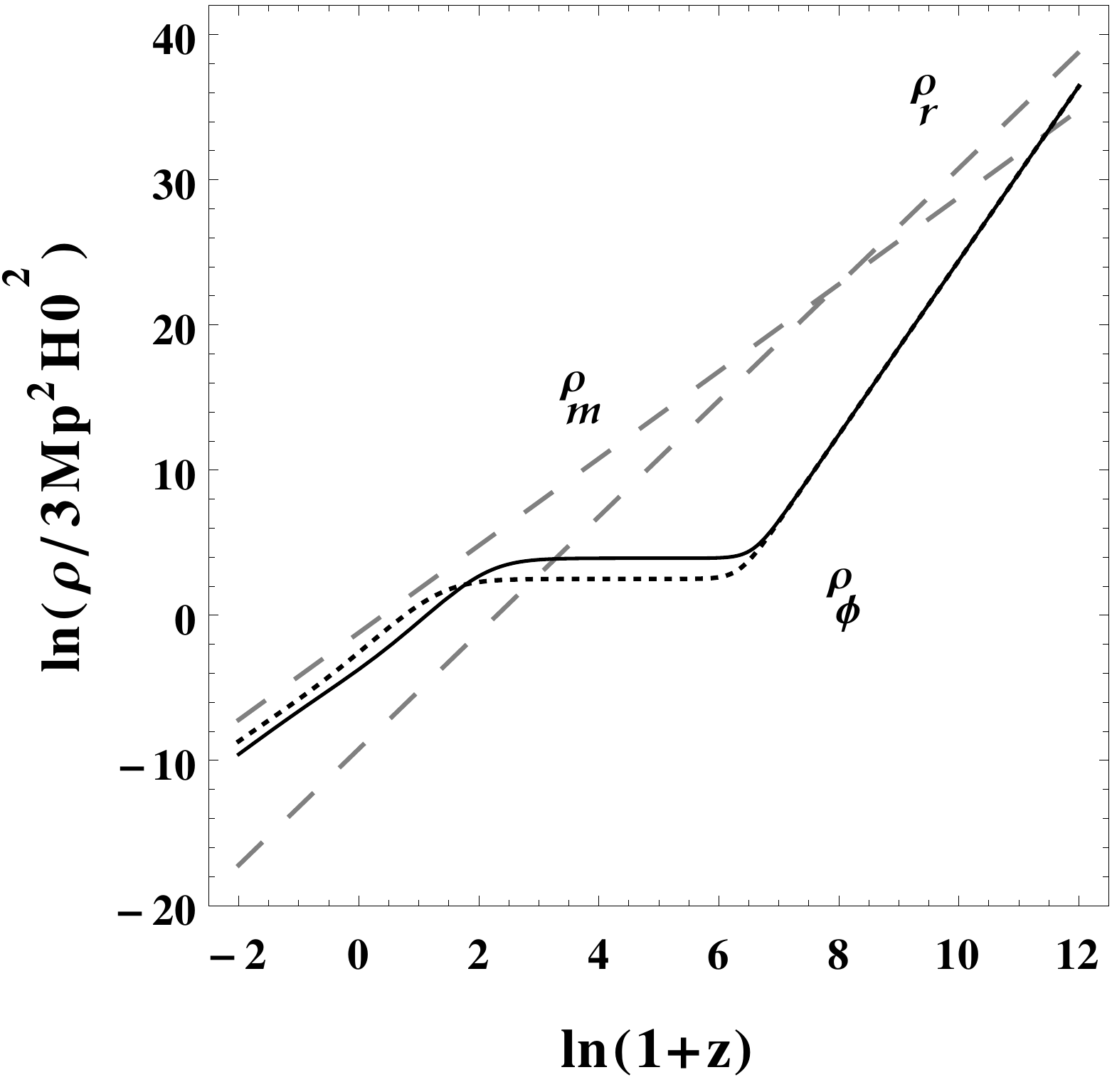}} 
\end{tabular}
\end{center}
\caption{ This figure exhibits the evolution of energy densities of field and background (matter and radiation; dashed lines) versus redshift. The field energy density is obtained for the potential (\ref{eq:pot}) with $n=1$ and $\alpha=4$ (dotted), 6 (solid line). In this case, field energy density scales with the background. Hence, we never get late-time acceleration.}
\label{fig:rhon1}
\end{figure}

In order to understand the late-time cosmological dynamics of the potential (\ref{eq:pot}), let us define the function $\Gamma$ as \cite{stein}:
\begin{eqnarray}
\Gamma=\frac{V (\phi) ~V(\phi)_{,\phi\phi}}{\left( V(\phi)_{,\phi}\right) ^2}
\label{eq:gammaV}
\end{eqnarray}
The properties of $\Gamma$ determine whether the tracking solutions exist or not. For any choice of the potential, $\Gamma$ could be $<$, $>$ or $=1$. These three conditions describe three different solutions such as $\Gamma < 1$ (thawing), $\Gamma = 1$ (scaling) and $\Gamma > 1$ (tracker) \cite{stein}. 

For potential (\ref{eq:pot}), the explicit form of $\Gamma$ can be written as 
\begin{eqnarray}
\Gamma= 1+ \frac{n-1}{n \alpha} ~\left( \frac{M_p}{\phi}\right)^n 
\label{eq:gamma}
\end{eqnarray}
From Eq. (\ref{eq:gamma}) one can notice that the function $\Gamma$ is unity for $n=1$ that corresponds to the standard exponential potential. The evolution of $\Gamma$ versus field for various values of $n$ is displayed in the right panel of Fig. \ref{fig:v}. From this figure, for $n=1$, one can clearly see that the evolution of $\Gamma$ is constant and remains unity throughout the evolution of $\phi$. In this case, the scalar field $\phi$ rolls down the potential from steep region, and its energy density undershoots the background, remains sub-dominant and scales as $\rho_\phi \sim a^{-6}$, where $a$ is the expansion factor of the universe. The evolution of the scalar field freezes due to the large Hubble damping, and its energy density becomes similar to the background, finally it exits from the freezing behavior and scales with the background upto the present epoch, and remains so in future. Hence there is no late-time cosmic acceleration. This kind of solutions are known as scaling solutions \cite{wand}. The scaling behavior of the field energy density for such an exponential potential is shown in Fig. \ref{fig:rhon1}.

For $n>1$ and large field values, the function $\Gamma$ approaches to unity this implies that the behavior of the potential  looks like to the standard exponential one, see Eq. (\ref{eq:gamma}) and right panel of Fig \ref{fig:v}. However, as $\phi$ approaches to the origin, the function $\Gamma$ deviates from unity, and later Fig. \ref{fig:v} confirm this. In order to get late-time cosmic acceleration one has to exit from scaling regime, and that's only possible when we consider more steeper potential than the exponential one. Therefore, in this letter, we shall study an exponential potential (\ref{eq:pot}) with $n>1$.

In a spatially flat Freidmann-Lemaitre-Robartson-Walker (FLRW) universe, the equations of motion have following forms:
\begin{eqnarray}
H^2 &=& \frac{1}{3 M_{p}^2}   \left( \rho_{\phi}+ \rho_m + \rho_r  \right)
\label{eq:H}
\end{eqnarray}
\begin{eqnarray}
\ddot{\phi} +3 H \dot{\phi}+ V'(\phi)=0
\label{eq:phidd}
\end{eqnarray}
where $H$ is the Hubble parameter. The quantities $\rho_{\phi}$, $\rho_{m}$ and $\rho_{r}$ represent the energy densities of a scalar field, matter and radiation, respectively. The dot and prime $(')$ represent the derivative with respect to cosmic time and scalar field, respectively. 

\begin{figure}[tbp]
\begin{center}
\begin{tabular}{ccc}
{\includegraphics[width=2.1in,height=2.1in,angle=0]{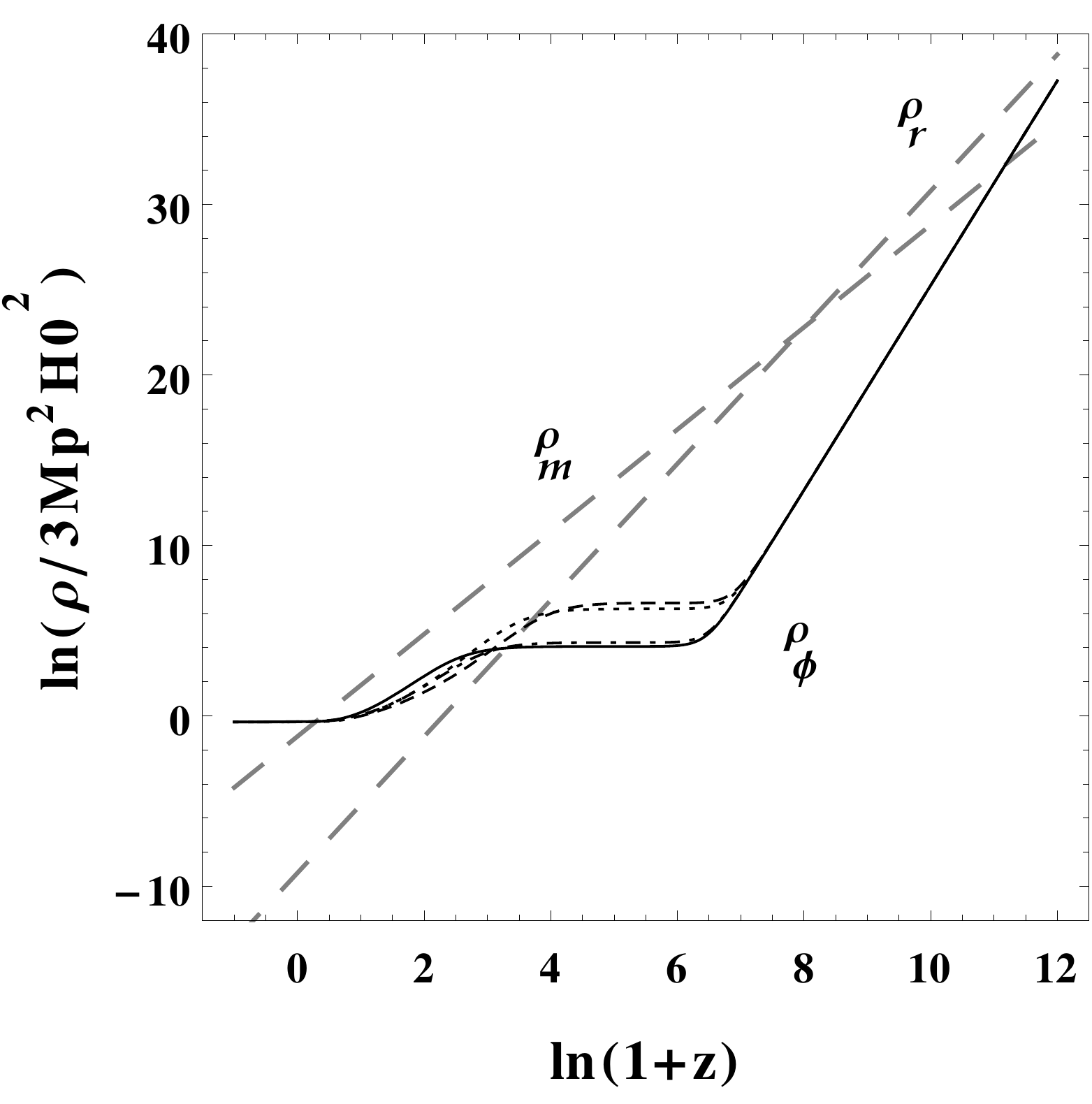}} &
{\includegraphics[width=2.1in,height=2.1in,angle=0]{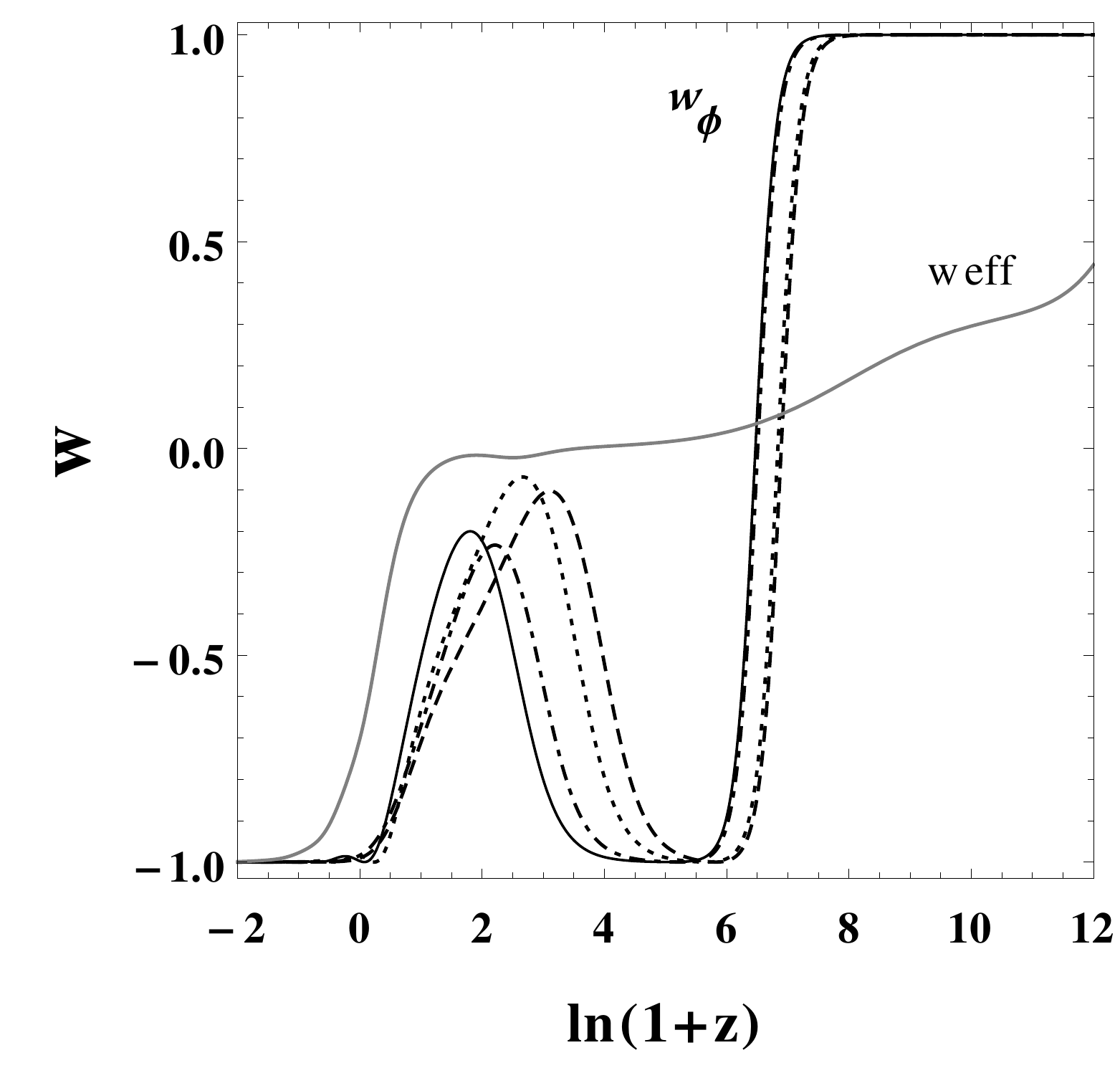}} &
{\includegraphics[width=2.1in,height=2.1in,angle=0]{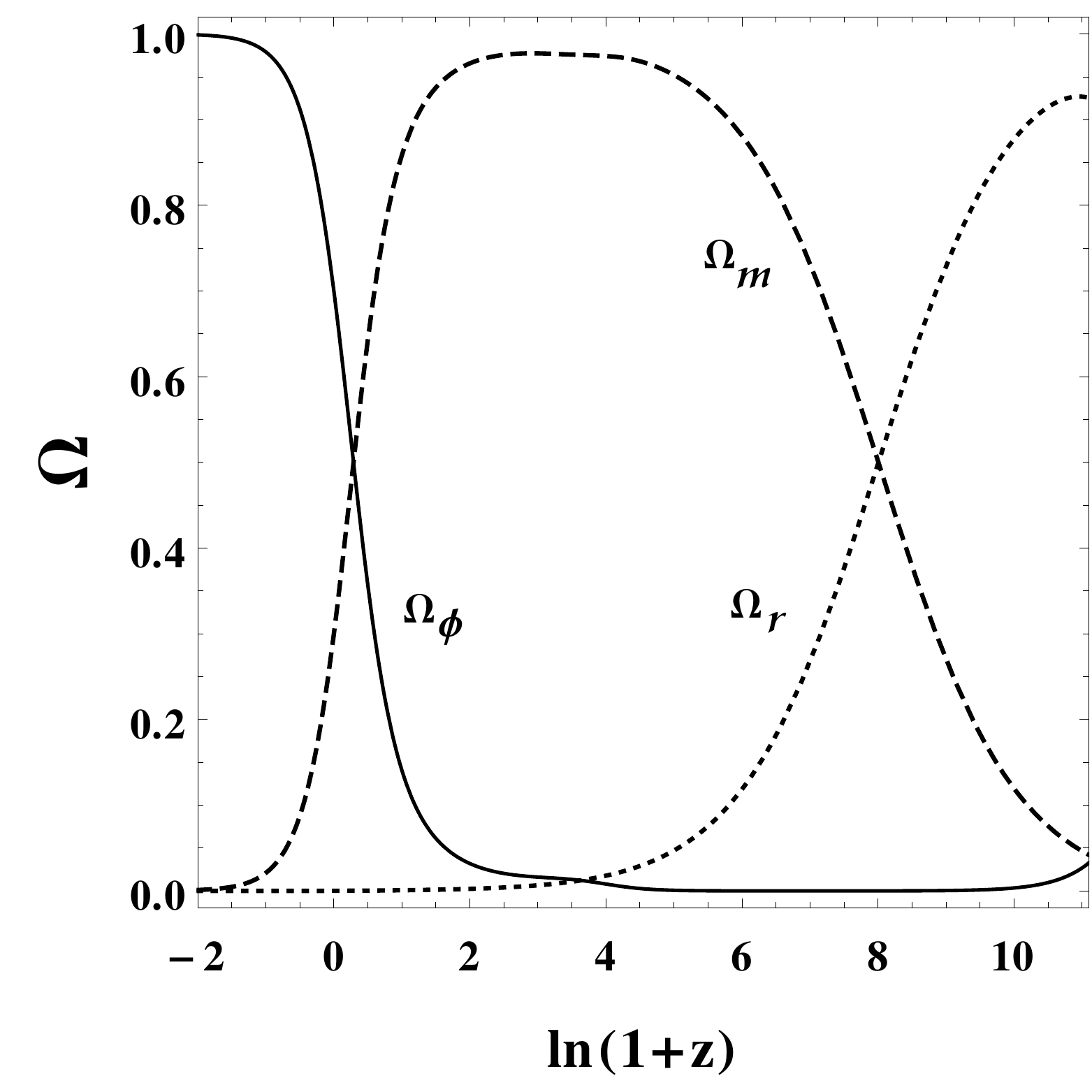}}
\end{tabular}
\end{center}
\caption{This figure corresponds to potential (\ref{eq:pot}) with $n>1$ and different values of $\alpha$. The left and middle panels show the evolution of $\rho_{\phi}$, $\rho_m$, $\rho_r$, $w(\phi)$ and $w_{eff}$ versus $z$. Both panels are obtained for $n=2$, $\alpha=4$ (solid), $n=2$, $\alpha=6$ (dotted), $n=$3, $\alpha=4$ (dot-dashed) and $n=$3, $\alpha=6$ (dashed). The right panel exhibits the evolution of energy density parameter $\Omega$ versus $z$ for $n=3$ and $\alpha=6$. The solid, dashed and dotted lines represent $\Omega_{\phi}$, $\Omega_m$ and $\Omega_r$, respectively. }
\label{fig:rhon2}
\end{figure}

The equation of state (EOS) $w(\phi)$, effective EOS $w_{eff}$, and the energy density parameters for the model under consideration are defined as
\begin{eqnarray}
\label{eq:w}
w(\phi)&=& \frac{p_{\phi}}{\rho_{\phi}}\\
w_{eff}&=& -1-\frac{2 \dot{H}}{3H^2}\\
\label{eq:weff}
\Omega_{\phi}&=& \frac{\rho_{\phi}}{\rho_c}, \qquad \Omega_{m}= \frac{\rho_{m}}{\rho_c}, \qquad  \Omega_{r}= \frac{\rho_{r}}{\rho_c}
\label{eq:omega}
\end{eqnarray}
where $\rho_{\phi}= \dot{\phi}^2/2+V(\phi)$ and $\rho_c=3M_p^2H^2$. In the discussion to follow let us consider the following dimensionless quantities
\begin{eqnarray}
Y_1={\phi \over M_p},\quad Y_2={\dot{\phi} \over M_{p} H_{0}},\quad {\cal V}={ V(Y_1) \over M_p^2 H_0^2}.
\label{eq:dimensionq}
\end{eqnarray}
which are used to form a system of first-order differential equations
\begin{eqnarray}
\label{eq:auto1}
\frac{dY_1}{dN} &=& \frac{ Y_2 H_0}{H(Y_1,Y_2)} \,\\
\frac{dY_2}{dN} &=& -3Y_2-{H_0 \over H(Y_1, Y_2)}\Big[{d {\cal V}(Y_1) \over dY_1} \Big]
\label{eq:auto2}
\end{eqnarray}
where $N=lna$, and the function $H(Y_1, Y_2)$ is given as 
\begin{eqnarray}
H(Y_1, Y_2) &=& H_0 \sqrt{\left[{Y_2^2 \over 6}+ {{\cal V}(Y_1) \over 3} +{\Omega_{0m} e^{-3a}} +{\Omega_{0r} e^{-4a}}  \right]}
\label{eq:h}
\end{eqnarray}
Here, $\Omega_{0r}$ and $\Omega_{0m}$ are the current energy density parameters of radiation and matter, respectively. The parameter $H_0$ is the current value of Hubble parameter. We numerically evolve the equations (\ref{eq:auto1}) and (\ref{eq:auto2}); the results are displayed in Figs. \ref{fig:rhon1}$-$\ref{fig:data}. Fig. \ref{fig:rhon1} shows the evolution of scalar field energy density versus redshift for an exponential potential with $n=1$. In this case, we do not obtain late-time acceleration as the field evolves from the steep region and approaches to the origin, the field energy density $\rho_\phi$ undershoots the background and freezes for a while due to a large Hubble damping. Around the present epoch, the field again starts evolving and its energy density mimics the background, and remain so in future.

Next, we consider the potential (\ref{eq:pot}) with $n>1$. Fig. \ref{fig:rhon2} exhibits the evolution of the field energy density $\rho_\phi$, EOS $w(\phi)$ and energy density parameter $\Omega$ versus redshift. Initially, $\rho_\phi$ undershoots the background and remains sub-dominant for most of the time of evolution as field does not evolve due to the increased value of Hubble damping. Around the present epoch, it switches over and converts to scaling behavior. At late-time, it exits from scaling behavior and derives the current accelerated expansion of the universe. This kind of solutions are known as tracker, and the behaviors are shown in Fig. \ref{fig:rhon2}.
\section{Statefinder and $Om$ diagnostics}
\label{sec:om}
In the literature, important geometrical diagnostics have been proposed, namely statefinder and $Om$ \cite{Sahni1,Sahni2,Z}. We shall use these diagnostics to distinguish DE models among themselves and with $\Lambda$CDM. Statefinders rely on the second and third order derivatives of the expansion factor with respect to time, whereas $Om$ depends only on the first order derivative. As a result, $Om$ is much simpler diagnostic when applied to observations. Both diagnostics have been extensively studied in the past few years to discriminate various models of DE \cite{alam1,alam2,alam3}. Following Ref. \cite{Sahni1}, we define statefinder pairs $\{r,s\}$ and $\{r,q\}$ as
\begin{equation}
q=-\frac{\ddot{a}}{aH^{2}}\text{, \ \ } r=\frac{\dddot{a}}{aH^{3}}\text{, \ \ }s=\frac{r-1}{3(q-\frac{1}{2})}\text{,}
\label{eq:rsq}
\end{equation}
where $q$ and $H$ represent the deceleration and Hubble parameters, respectively. The parameter ``a" denotes the expansion factor of the universe, and dot designates the derivative with respect to cosmic time.

In a spatially flat FLRW background, the statefinder pairs in terms of EOS $(w)$ can be written as \cite{Sahni1}:
\begin{eqnarray}
\label{eq:q_w}
q &=& \frac{1}{2} \left(1+3w \Omega_X \right),\\
\label{eq:r_w}
r &=& 1+ \frac{9w}{2} \Omega_X (1+w)-\frac{3}{2} \Omega_X \frac{\dot{w}}{H},\\
s &=& 1+w- \frac{\dot{w}}{3wH},
\label{eq:s_w}
\end{eqnarray}
where $\Omega_X=1-\Omega_m$; $\Omega_m$ is the energy density parameter of matter. From Eqs. (\ref{eq:q_w}), (\ref{eq:r_w}) and (\ref{eq:s_w}), we conclude following:
\begin{eqnarray}
\text{For}~ \Lambda \text{CDM}~ (w=-1), && r=1~~ \text{and}~~ s=0,\nonumber\\
\text{for SCDM}~ (w=0), && r=1~~ \text{and}~~ q=1/2,\nonumber\\
\text{for dS,} && r=1~~ \text{and}~~ q=-1,
\label{eq:qrs2}
\end{eqnarray}
where SCDM and dS stand for standard cold dark matter and de-Sitter expansion of the universe.
\begin{figure}[tbp]
\begin{center}
\begin{tabular}{ccc}
{\includegraphics[width=2.1in,height=2.1in,angle=0]{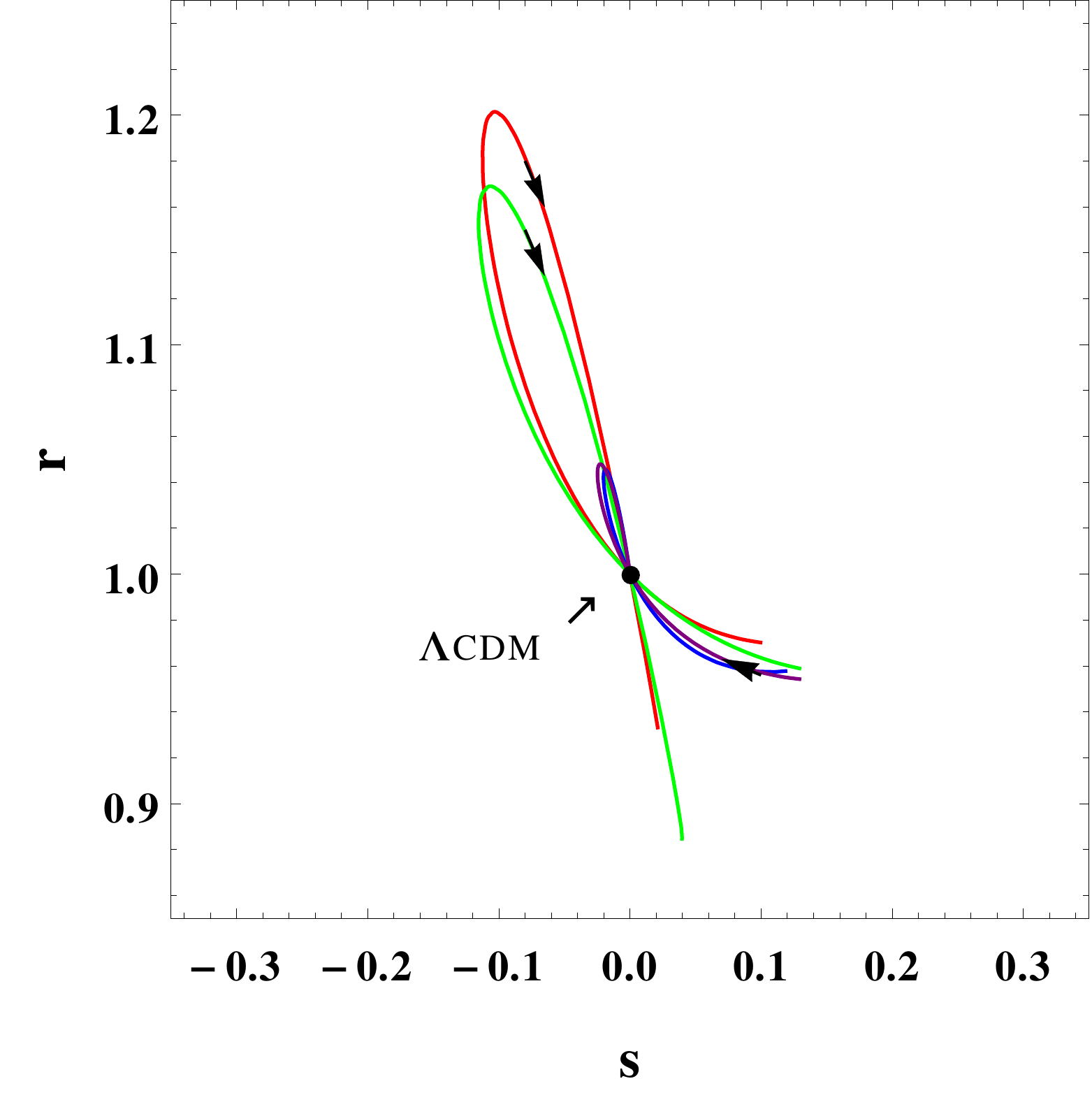}} & 
{\includegraphics[width=2.1in,height=2.1in,angle=0]{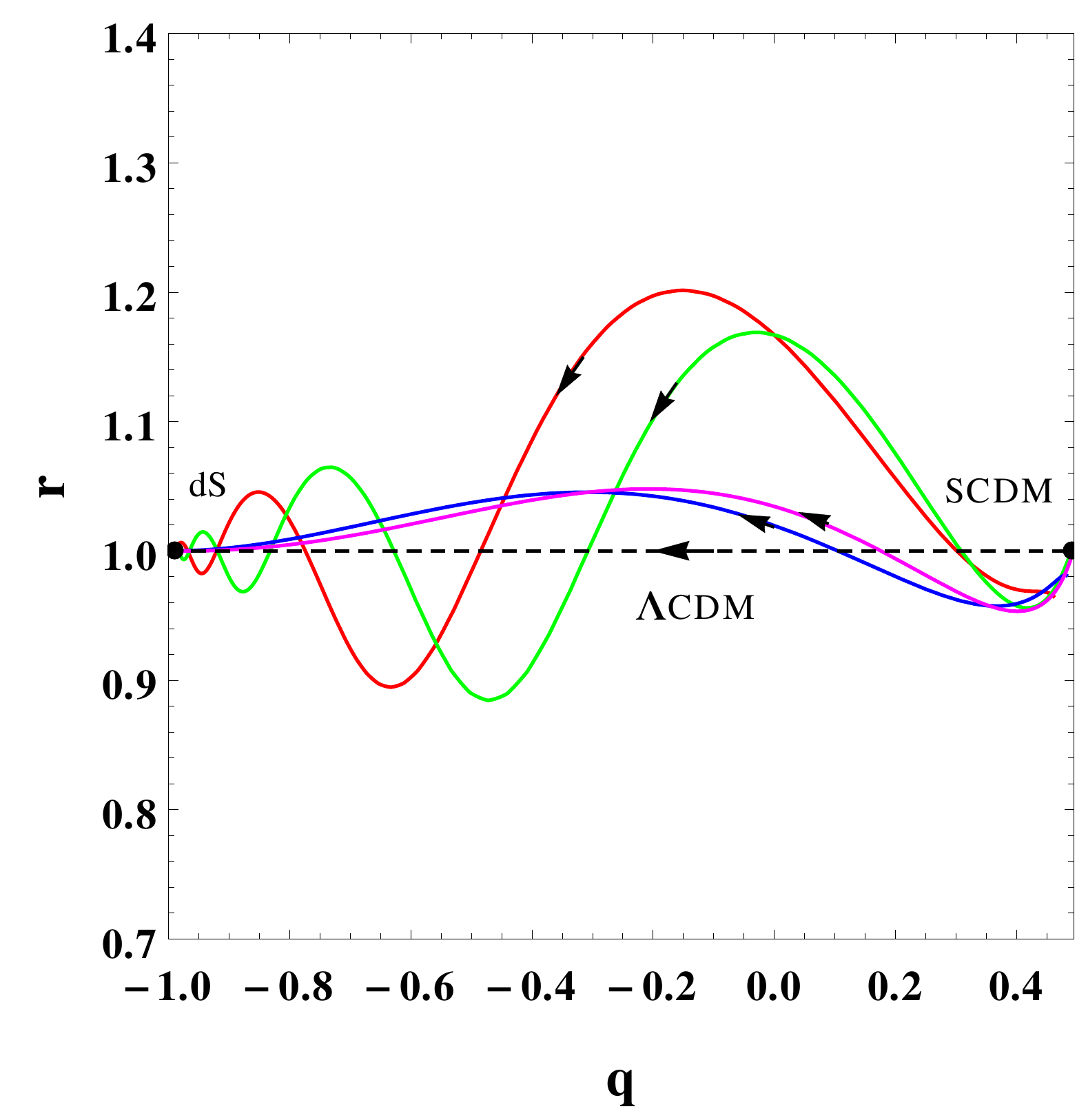}} &
{\includegraphics[width=2.1in,height=2.1in,angle=0]{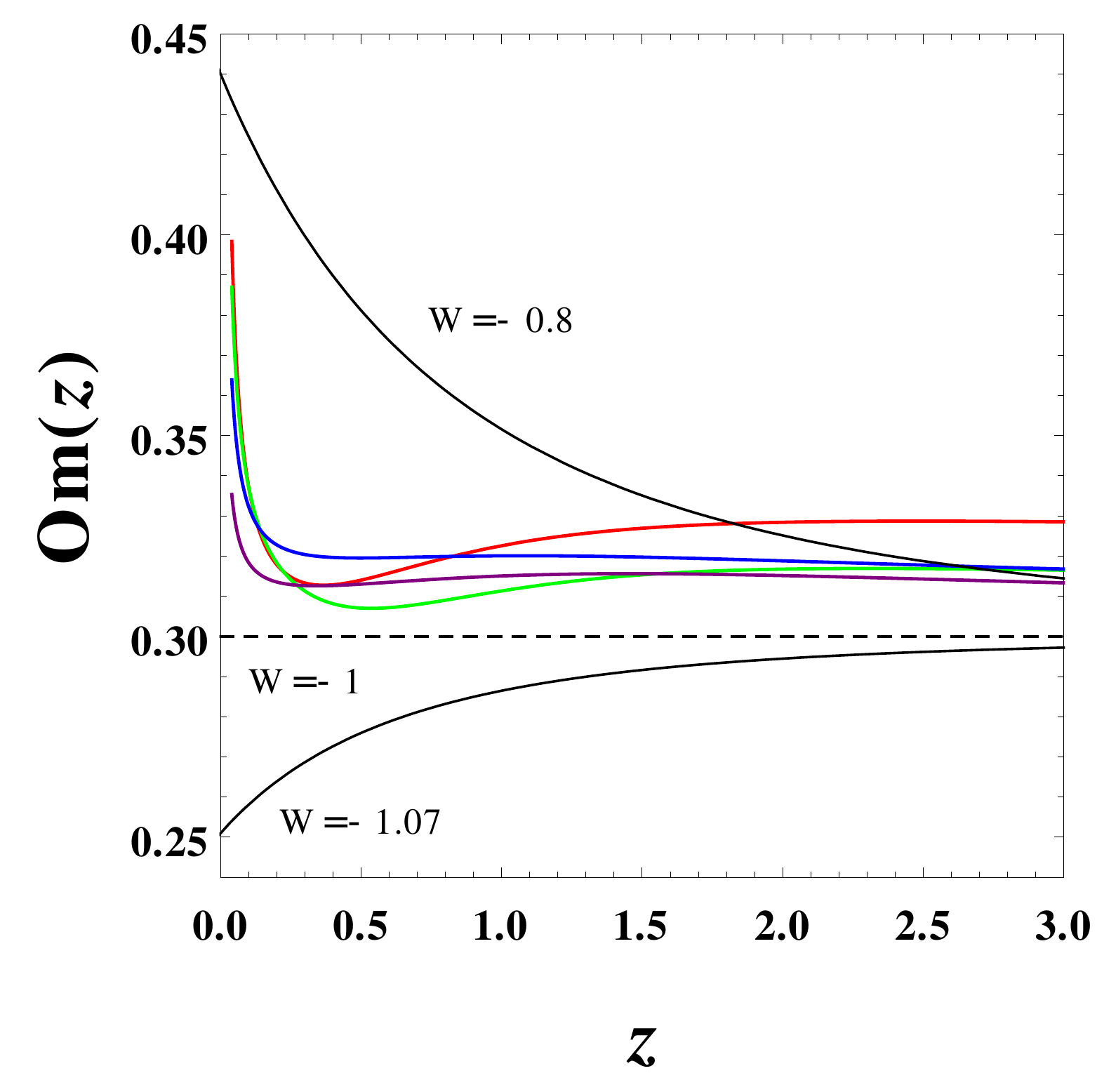}}
\end{tabular}
\end{center}
\caption{This figure displays the evolution of statefinder pairs $\{r,s\}$, $\{r,q\}$ and $Om(z)$ for $n>1$ and different values of $\alpha$ as Red (n=2 , $\alpha=4$), Green (n=2 , $\alpha=6$), Blue (n=3 , $\alpha=4$) and Purple (n=3 , $\alpha=6$). In the left panel, fixed point $(r=1,s=0)$ denotes $\Lambda$CDM. All trajectories in $r-s$ plane pass through $\Lambda$CDM. In the middle panel, the fixed points $(r=1,q=0.5)$ and $(r=1,q=-1)$ represent SCDM and dS. All trajectories in $r-q$ plane diverge from SCDM and converge to dS. The horizontal dashed line corresponds to $\Lambda$CDM. In the right panel, we show the evolution of $Om(z)$ versus $z$ for $\Lambda$CDM $(w=-1)$, quintessence $(w=-0.8)$ and phantom $(w=-1.07)$, and also for potential (\ref{eq:pot}) with $n>1$. The evolution of $Om(z)$ for $n>1$ has negative curvature which is compatible with the analytical solution (\ref{eq:om2}). }
\label{fig:rs}
\end{figure}

For the potential (\ref{eq:pot}) with $n=2$ and 3, we obtain different trajectories in the $r-s$ and $r-q$ planes, and study their behaviors. The evolution of statefinders for different values of $n$ and $\alpha$ are shown in the left and middle panels of Fig. \ref{fig:rs}. In the left panel, the fixed point $(r=1, s=0)$ corresponds to $\Lambda$CDM. All the trajectories pass through $\Lambda$CDM. The middle panel of Fig. \ref{fig:rs} exhibits the evolution of $r$ versus $q$. In this panel, all trajectories diverge from the fixed point $(r=1, q=0.5)$ that corresponds to SCDM, and converge to another point $(r=1, q=-1)$ that represents dS expansion. The horizontal dashed line depicts $\Lambda$CDM behavior. The models in which late-time acceleration might arise without DE do not converge to dS due to their phantom behavior in future \cite{alam2017}.

We mentioned that $Om$ is a simpler diagnostic when applied to observations as it explicitly depends on the Hubble parameter and redshift. In a spatially flat FLRW universe, it is defined as \cite{Sahni2,Z}:
\begin{equation}
Om\left( z\right) =\frac{\left( H(z)/H_{0}\right)^{2}-1}{\left( 1+z\right)
^{3}-1}\text{} 
\label{eq:om}
\end{equation}
For a spatially flat universe, the Hubble parameter with constant EOS is given by
\begin{equation}
H^2(z) = H_0^2 \left[ \Omega_{0m}(1+z)^3 + (1-\Omega_{0m})(1+z)^{3(1+w)}\right] ,
\label{eq:Hconst}
\end{equation}
The corresponding expression of $Om(z)$  for constant EOS is written as
\begin{equation}
Om(z) = \Omega_{0m} + (1-\Omega_{0m})\frac{(1+z)^{3(1+w)}-
1}{(1+z)^3-1}
\label{eq:omconst}
\end{equation}
From Eq. (\ref{eq:omconst}), one can clearly see that
\begin{eqnarray}
Om(z) &=& \Omega_{0m} \qquad \text{for}~ \Lambda \text{CDM}~ (w=-1)\nonumber\\
Om(z) &>& \Omega_{0m} \qquad \text{for quintessence}~ (w>-1)\nonumber\\
\text{and}~~ Om(z) &<& \Omega_{0m} \qquad \text{for phantom}~ (w<-1)
\label{eq:om2}
\end{eqnarray} 
The evolutions of $Om(z)$ for $\Lambda$CDM, quintessence and phantom are displayed in the right panel of Fig. \ref{fig:rs}. From this figure, one can notice that $\Lambda$CDM $(w=-1)$, quintessence $(w=-0.8)$ and phantom $(w=-1.07)$ have zero, negative and positive curvatures, respectively. In Fig. \ref{fig:rs}, we also exhibit the evolution of $Om(z)$ for potential (\ref{eq:pot}) with different values of $n$ and $\alpha$. The evolution of $Om(z)$ for the underlying models show negative curvatures which are consistent with the analytical solution, see Eq. (\ref{eq:om2}). Although, the models in which late-time acceleration might arise due to the coupling between baryonic matter and dark matter without the presence of extra degrees of freedom have positive curvatures however $w>-1$ at the present era. More precisely, $Om(z)$ has positive curvature for the models that provide late-time acceleration without DE though they have $w>-1$ at the current epoch \cite{alam2017}.
\section{Observational Constraints}
\label{sec:data}
\begin{figure}[tbp]
\begin{center}
\begin{tabular}{c}
{\includegraphics[width=5.1in,height=4.1in,angle=0]{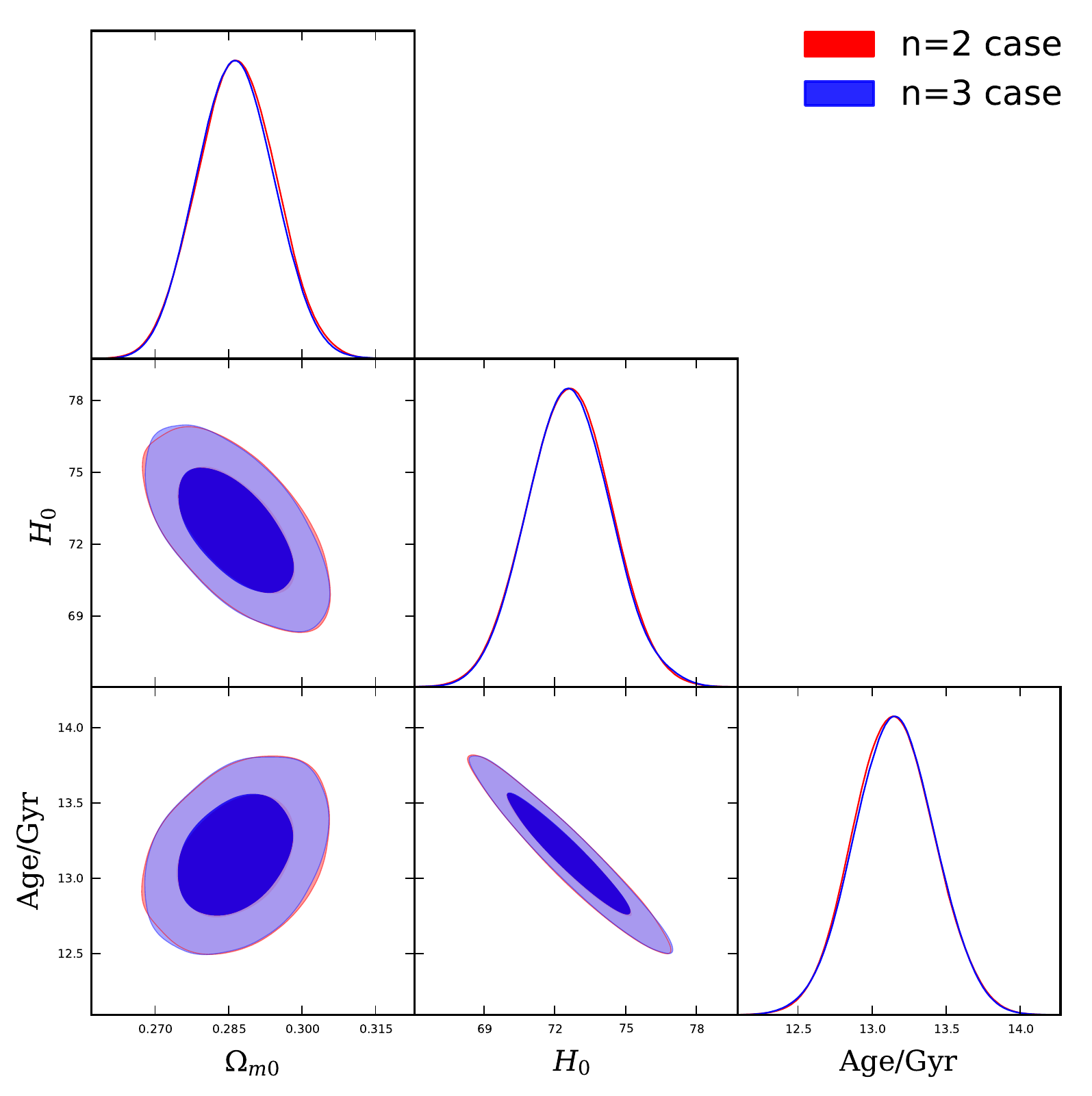}} 
\end{tabular}
\end{center}
\caption{This figure shows 68\% and 95\% confidence-level contour plots for the parameters $\Omega_{m0}$, $H_0$ and derived parameter Age as well as their one dimensional marginalized distribution of individual parameters. The figure is displayed for $n=2$ and 3 with the combined analysis of BAO, JLA and HST.}
\label{fig:data}
\end{figure}

In this section, we briefly describe the astronomical data and corresponding methodology that have been used to constrain the scalar field models of DE.
\begin{itemize}
\item \textbf{BAO data:} The baryon acoustic oscillation data are powerful to probe the nature of dark energy. In this analysis, we use four BAO points: the 6dF Galaxy Survey (6dFGS) measurement at $z_{\emph{\emph{eff}}}=0.106$ \cite{ref:BAO1-Beutler2011}, the Main Galaxy Sample of Data Release 7 of Sloan Digital Sky Survey (SDSS-MGS) at $z_{\emph{\emph{eff}}}=0.15$ \cite{ref:BAO2-Ross2015}, CMASS and LOWZ samples from the latest Data Release 12 (DR12) of the Baryon Oscillation Spectroscopic Survey (BOSS) at $z_{\mathrm{eff}}=0.32$ \cite{ref:BAO3-Gil-Marn2015} and $z_{\mathrm{eff}}=0.57$ \cite{ref:BAO3-Gil-Marn2015}.

\item \textbf{JLA data:} Type Ia supernovae provide the first signal for an accelerating universe, and still they serve as the main observational data to probe the late-time acceleration of the universe. In the current analysis, we use latest compilation of the SNIa, namely JLA sample \cite{ref:JLA} that contains 740 SNIa data points in the redshift range $z\in[0.01, 1.30]$.

\item \textbf{HST data:} According to the Hubble Space Telescope probe, we include the local value of Hubble constant as $H_0= 73.02 \pm 1.79$ km/s/Mpc which is obtained with 2.4\% precision by the Riess et al. \cite{ref:Riess2016}.
\end{itemize}
\begingroup
\begin{center}
\begin{table}
\caption{This table summarizes the constraint results on the parameters $\alpha$, $\Omega_{m0}$, $H_0$ and derived parameter Age of scalar field DE model with $n=2$ using the joint analysis of BAO, JLA and HST.}
\begin{tabular}{ccc}
\hline\hline
Parameters~~~~ & Mean with errors~~~~ & Best-fit value\\ \hline\\
$\alpha$&$5.04_{-5.04-5.04}^{+4.96+4.96}$&$8.87$\\\\
$\Omega_{m0}$&$0.287_{-0.008-0.015}^{+0.008+0.0156}$&$0.286$\\\\
$H_0$&$72.6_{-1.8-3.4}^{+1.7+3.5}$&$72.6$\\\\
${\rm{Age}}/{\rm{Gyr}}$&$13.1_{-0.3-0.5}^{+0.3+0.5}$&$13.2$\\\\
\hline\hline
\end{tabular}
\label{tab:results1}
\end{table}
\end{center}
\endgroup
\vskip0.5mm
\begingroup
\begin{center}
\begin{table}
\caption{This table shows the constraint results on the parameters of DE model with $n=3$ using an integrated data base of BAO, JLA and HST.}
\begin{tabular}{ccc}
\hline\hline
Parameters~~~~ & Mean with errors~~~~ & Best-fit value \\ \hline\\
$\alpha$&$5.01_{-5.01-5.01}^{+4.99+4.99}$&$3.7$\\\\
$\Omega_{m0}$&$0.286_{-0.008-0.015}^{+0.008+0.015}$&$0.285$\\\\
$H_0$&$72.6_{-1.8-3.4}^{+1.8+3.5}$&$72.9$\\\\
${\rm{Age}}/{\rm{Gyr}}$&$13.2_{-0.3-0.5}^{+0.3+0.5}$&$13.1$\\\\
\hline\hline
\end{tabular}
\label{tab:results2}
\end{table}
\end{center}
\endgroup
To put the constraints on the model parameters, we use the likelihood as $\mathcal{L}\propto e^{-\chi^2_{tot}/2}$. Here, $\chi^2_{tot} = \sum _{i} \chi^2_i$, and $i$ runs over the all data sets that we use that is BAO, JLA and HST. We modify the publicly available code CosmoMC \cite{ref:cosmomc-Lewis2002,ref:camb}; a Markov Chain Monte Carlo simulation to extract the cosmological parameters associated with the models. The potential (\ref{eq:pot}) has two particular parameters $n$ and $\alpha$. In our analysis, we talk about the fixed $n$ with $n=2$ and 3 cases, for the models under consideration. Therefore, we have following three parameter space given as
\begin{equation}
P=\{H_0, \Omega_{m0}, \alpha\}
\end{equation}
In the data-fitting, the priors are
\begin{equation}
H_0\in[20,100], \qquad \Omega_{m0}\in[0.001,0.99] \qquad \text{and} \qquad \alpha\in[0,10]
\end{equation}
In Tables \ref{tab:results1} and \ref{tab:results2}, we summarize our results for the underlying models with $n=2$ and 3 using the joint data sets of BAO, JLA and HST. Fig. \ref{fig:data} shows the $68\%$ and $95\%$ confidence-level contour plots for different pairs of parameters such as $\Omega_{m0}$, $H_0$ and derived parameter Age as well as their one dimensional marginalized distribution of individual parameters. From our analysis, we find that the constraint results are almost same for the cases $n=2$ and 3 as the best-fit likelihood are found to be $\chi^2_{min}=703.57$ $(n=2)$ and $\chi^2_{min}=703.60$ $(n=3)$. For $n=2$ and 3, the parameters $H_0$ and $\Omega_{m0}$ are found to be tightly constraint in a combined analysis of BAO, JLA and HST, whereas the constraint results of parameter $\alpha$ are divergent, see Tables \ref{tab:results1} and \ref{tab:results2}.
\section{Conclusion}
\label{sec:conc}
In this paper, we have investigated the scalar field models of DE based upon the steeper potentials than exponential. This kind of potentials have remarkable properties. For $n=1$, the function $\Gamma$ is unity throughout the evolution, see Fig. \ref{fig:v}. In this case, field energy density $\rho_{\phi}$ scales with the background and does not provide late-time acceleration that is shown in Fig. \ref{fig:rhon1}. This kind of solution is known as scaling solution. In case of $n>1$ and large field values, $\Gamma$ approaches to unity and the behavior is similar to the standard exponential potential. However, for small field values i.e. as $\phi$ approaches to origin, $\Gamma$ shows deviation from unity $(\Gamma > 1)$ which is displayed in Fig. \ref{fig:v}. Such solutions are known as tracker and derive late-time acceleration. In Fig. \ref{fig:rhon2}, we have shown the evolution of energy density $\rho_{\phi}$, EOS and energy density parameter for $n>1$ and different values of $\alpha$. Initially, the field evolves from steep region and moves towards the origin, energy density $\rho_{\phi}$ undershoots the background and freezes for a while due to large Hubble damping. As time passes, further field evolves, its energy density becomes comparable to the background energy density, and follows scaling behavior around the present epoch. At late-time, it exits from scaling regime and gives late-time acceleration, see Fig. \ref{fig:rhon2}. To the best of our knowledge, this characteristic of the steeper potential (\ref{eq:pot}) was not discussed earlier in the literature.

In addition, we used statefinder and $Om$ diagnostics to distinguish the underlying models among themselves and with $\Lambda$CDM. We showed the evolutions of $\{r,s\}$, $\{r,q\}$ and $Om$ for various values of $n$ and $\alpha$. In the $r-s$ plane, the fixed point $(r=1, s=0)$ corresponds to $\Lambda$CDM, and all trajectories pass through this point. In the $r-q$ plane, $\Lambda$CDM and different trajectories for various values of $n$ and $\alpha$ diverge from the fixed point $(r=1, q=0.5)$ that represents SCDM and converge to $(r=1, q=-1)$ that corresponds to dS expansion. We have shown that our models are tracker that fall into quintessence class. For quintessence, the evolutions of $Om(z)$ for different values of $n$ and $\alpha$ have negative slope that is consistent with the analytical solution (\ref{eq:om2}), see right panel of Fig. \ref{fig:rs}.

To put the observational constraints on the model parameters, we modified the publicly available CosmoMC code \cite{ref:cosmomc-Lewis2002,ref:camb} and used an integrated data base of BAO, SNIa from JLA sample and HST. The best-fit values of the model parameters for $n=2$ and 3 cases are shown in Tables \ref{tab:results1} and \ref{tab:results2}. The 68\% and 95\% confidence-level contour plots for the parameters $\Omega_{m0}$, $H_0$ and derived parameter Age as well as their one dimensional marginalized distribution of individual parameters are displayed in Fig. \ref{fig:data}.
\section*{Acknowledgements}
M.S. thanks M. Sami and Qiang Wu for fruitful discussions. W. Yang's work is supported by the NNSFC Grants No. 11705079 and No. 11647153, China. A.W. acknowledges financial support provided by the NNSFC Grants No. 11375153 and No. 11675145, China.

\end{document}